 \documentclass[11pt,a4paper]{article}
\usepackage{epsfig}

\begin{document}
\renewcommand{\thefootnote}{\fnsymbol{footnote}} 

  \title{\emph{Ab initio} Simulations of Fe-based Ferric Wheels}
 \author{
 A.~V. Postnikov$^1$\footnotemark[1], Jens Kortus$^2$\footnotemark[2]
 and Stefan Bl\"ugel$^{1,3}$
\\*[2mm]
$^1$\emph{Universit\"at Osnabr\"uck -- Fachbereich Physik,}\\
    \emph{D-49069 Osnabr\"uck, Germany} \\
$^2$\emph{Max-Planck Institut f\"ur Festk\"orperforschung, Heisenbergstr. 1,}\\
    \emph{D-70569 Stuttgart, Germany} \\
$^3$\emph{Institut f\"ur Festk\"orperforschung, Forschungszentrum J\"ulich,}\\  
    \emph{D-52425 J\"ulich, Germany} 
        }  
\date{}  
\maketitle
\footnotetext[1]{Corresponding author. Tel. +49-541-969-2377, fax: -2351; 
E-mail: apostnik@uos.de.\\
Permanent address: Institute of Metal Physics, 620219 Yekaterinburg, Russia
                }  
\footnotetext[2]{Tel. +49-711-689-1664, fax: -1632; 
E-mail: j.kortus@fkf.mpg.de.
                }  

\begin{abstract}
Based on first-principles density-functional theory calculations we investigate 
the electronic structure of hexanuclear ``ferric wheels'' 
$M$Fe$_6$[N(CH$_2$CH$_2$O)$_3$]$_6$Cl ($M$ = Li, Na) 
in their antiferromagnetic ground state. The electronic structure 
is presented in form of spin- and site-resolved local densities of states. 
The latter clearly indicate that the magnetic moment 
is distributed over several sites. The local moment at the iron site is still 
the largest one with about 4 $\mu_{\mbox{\tiny B}}$, thus indicating  
the valence state of iron to be closer to Fe(II) than to commonly
accepted Fe(III). The local spin of $S$=5/2 per iron site, following from 
magnetization measurements, is perfectly reproduced if one takes the moments 
on the neighbor atoms into account. The largest magnetic polarization 
is found on the apical oxygen atom, followed by nitrogen bridging oxygens. 
These findings are confirmed by a map of spatial spin density. A further 
goal of the present study has been a comparative test of two different DFT 
implementations, {\sc Siesta} and NRLMOL. They yield a very good 
agreement down to small details in the electronic structure.
\end{abstract}

\section*{Introduction}

There has been a remarkable increase in the number of studies 
on molecular magnets during the last decade which
clearly is driven by the progress in the chemistry of such systems. 
Many different families of molecular magnetic systems composed of 
all possible transition-metal atoms in various combinations with
stabilizing organic ligands have been synthesized, their crystal structure 
determined and their magnetic properties analyzed \cite{Mol_Magnets}. 

While there is a large amount of experimental data available,
the theoretical description in terms of quantitative -- and often
also qualitative -- understanding is missing in some cases, in others it
stays behind the experimental progress. The traditional approach 
to describe these systems is based on model Hamiltonians 
(``exact'' diagonalization  within,  e.g., Heisenberg model).
One the other hand, with the advance in computer technology and the 
development in theoretical methods, most notably in density-functional theory 
(DFT) \cite{RMP71-1253}, first-principles calculations became feasible.
Both approaches simplify real systems and are prone to certain
shortcomings. The model Hamiltonians neglect the true chemical
environment and bonding. All interactions are reduced to just few model
parameters. One is dependent on experimental input in order to
estimate these parameters, and the accuracy of quantitative predictions
depends on the parameters chosen. Even more problematic, it is not 
\emph{a priori} clear which interactions are important and should be 
included in the model and which are negligible. 
In contrast, first-principles calculations based on quantum mechanics
use the atomic positions as the only experimental input and, as
chemical bonding is included in a straightforward way, they are able to provide 
microscopic foundations of the magnetic behavior. Unfortunately, the solution 
of Schr{\"o}dinger's equation is limited to a few electrons so that one has 
to use approximations. The chemical complexity of molecular magnets 
(or, their large number of atoms) excluded so far the 
``straightforward'' simulations using approaches of superior accuracy, 
like quantum Monte Carlo. Density-functional based 
calculations have become the working horse in today's first-principles 
simulations.  DFT has been proven to be very accurate and reliable in many 
cases, although electron correlation effects beyond the mean-field 
approximation in DFT are neglected, which limits the accuracy and may lead 
to failures in the description of a given system.

Both approaches, the model-based and DFT-based, are useful and important 
for our understanding, and are best used together. While DFT can provide 
the input parameters for models (e.g. the exchange parameters $J$ 
for a Heisenberg model),
the model Hamiltonians can deliver information on many-body effects,
temperature dependence, and quantum fluctuations,
which are beyond the reach of DFT. By describing the relevant physics 
from different points of view
one develops a complete picture where essential features do not remain
unexplored.

Concentrating in the following on DFT-based strategies, we note that
molecular magnets turn out to be hard cases for computational treatment 
-- for different reasons in different established calculation approaches.
The openness of the crystal structure creates problems for traditional
muffin-tin- or atomic sphere approximation-based tools, 
and requires requires a ``full-potential'' formalism instead.
The presence of transition metals and/or other constituents with deep
semicore states, which can not be neglected, creates problems for 
pseudopotential planewave methods. Moreover, the need for a large 
(albeit largely empty) ``simulation box'' around a molecule makes 
planewave-related methods expensive. An abundance of hydrogen and hence 
strong disbalance in atom ``sizes'' makes the linearized augmented plane wave 
scheme numerically demanding. The often unusual chemical composition and 
coordination reduces the usefulness of \emph{a priori} tuned basis sets 
in those tight-binding-type methods which heavily depend on pre-adjusted 
basis functions for their efficiency. 
Further, the option to treat non-collinear spin-density distribution 
and to have accurate forces on atoms  for studying structure relaxation 
and dynamics may be advantageous and sometimes necessary.
As a consequence, flexible, efficient and accurate methods (and codes) are
in demand, of which several have been successfully applied
to systems of reasonable complexity. 

\section*{Calculations}

As an example of a system of practical interest we consider the
hexanuclear ``ferric wheels'' $M$Fe$_6$[N(CH$_2$CH$_2$O)$_3$]$_6$Cl
($M$ = Li, Na, see Fig.~1), synthesized at the 
Institut f\"ur Organische Chemie in Erlangen \cite{AnChIE36-2482} 
and labeled as substances {\bf 4} and {\bf 3} in the latter publication.
These materials belong to a larger family of Fe-coronates, of which
8- and 10-member rings are also known. A common feature of these
substances is the bonding of Fe atoms via doubled oxygen bridges,
somehow resembling the 90$^{\circ}$ coupling of magnetic atoms
in transition-metal oxides. In the vicinity of each Fe atom
one finds a nitrogen atom, connected via C$_2$H$_4$ chains to oxygen atoms. 
The Li- and Na-centered molecules have a very similar structure, with only 
slightly different Fe--O--Fe angles (101.1$^\circ$ for Li-wheel
and 103.3$^\circ$ for Na-wheel).

\begin{figure}
\centerline{\epsfig{figure=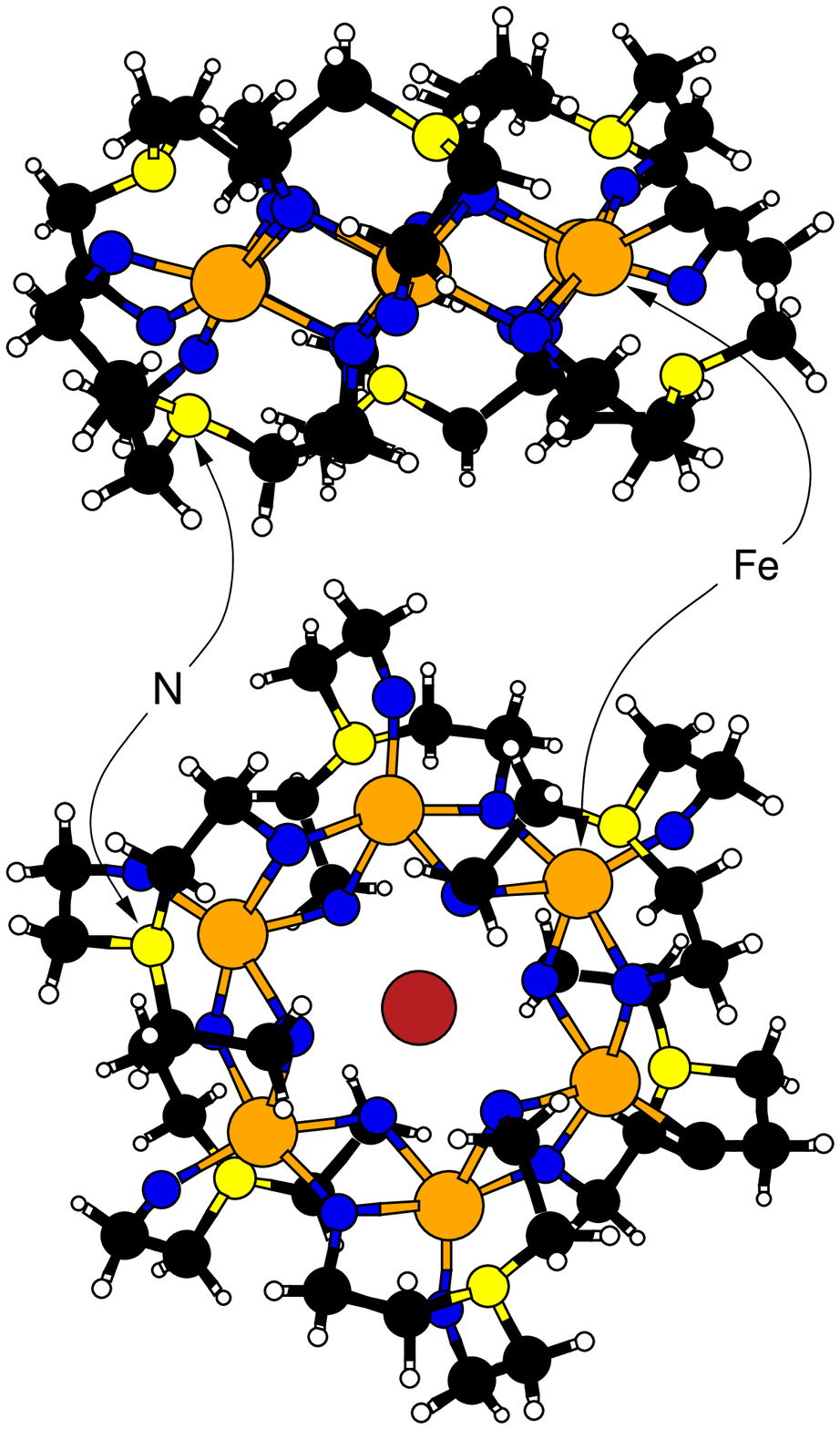,width=9.5cm}}
\bigskip
\caption{The Li-centered ``ferric wheel'' molecule. The Li ions
in the middle of the ring; the distant Cl ion included in the simulation
is not shown; the rest of (electrically neutral) solvent is neglected.}
\end{figure}

The interest in the magnetic properties of the system is 
the presence of large magnetic moments, traditionally argued
to correspond to $S$=5/2 on the Fe site, thus implying
a highly ionized Fe(III) state. Such an identification of the iron magnetic
moment is consistent with magnetization and torque measurements 
\cite{InCh38-5879}, from which Waldmann \emph{et al.} derived 
the values of exchange parameters $J$ for the spin Hamiltonian
of the Heisenberg model:
\begin{equation}
H = -J \left(\sum_{i=1}^5 {\bf S}_i{\cdot}{\bf S}_{i+1}
+ {\bf S}_6{\cdot}{\bf S}_1 \right) +
(\mbox{anisotropy term}) + (\mbox{Zeeman term})\,.
\end{equation}
The obtained values of $J$ are $-$18 to $-$20 K for the Li-wheel 
(depending on sample and method) and $-$22.5 to $-$25 K for the Na-wheel,
thus implying an antiferromagnetic (AFM) ground state. 
While the data from bulk magnetic measurements seem to be well  
established by now and reliable, virtually no spatially resolved
probing of the magnetic structure has been carried out yet. 
Recent X-ray photoelectron and X-ray emission spectroscopy studies 
\cite{our_XPS} allow to probe the electronic structure in the valence band 
and on the Fe site, albeit without resolution in spin. This is exactly 
where \emph{ab initio} calculations may yield a valuable contribution 
in providing site- and spin-resolved densities of states, as well as 
the spatial distribution of charge and magnetization densities. 
The verification of these predictions demands the comparison with 
(atom- and/or spin-averaged) spectroscopic data, and eventually comparison 
with other calculational approaches  at a different level of approximations.

In the present study, we compare the results of electronic structure
calculations by two different methods -- {\sc Siesta}\cite{JPCM14-2745}
and NRLMOL \cite{PRB41-7453}. 
Both methods are based on the DFT; from the various exchange-correlation
functionals implemented we used the generalized gradient approximation after 
Perdew--Burke--Ernzerhof \cite{PRL77-3865} for the present study. 
A further similarity is that their basis sets are atom-centered functions -- 
numerical pseudoatomic orbitals in {\sc Siesta} 
\cite{PRB64-235111} 
and contracted Gaussian-type orbitals in NRLMOL \cite{PRA60-2840}.
The main difference is that {\sc Siesta} uses norm-conserving pseudopotentials 
whereas NRLMOL is an all-electron code. NRLMOL uses the symmetry, which is 
relatively high in case of the Fe$_6$-wheels, of a molecule in question 
in all numerical tasks. Moreover, all computationally intensive parts 
of the code are massively parallelized \cite{PSSB217-197}, which makes DFT 
calculations on molecular magnets (with a typical size of 100--200 atoms) 
feasible. {\sc Siesta} is competitive due to the compactness and efficiency
of its numerical basis sets. Moreover, it can work as an order-$N$ method
(albeit not yet for magnetic systems), solving the DFT problem without 
explicit diagonalization. 
Other features of the two calculational methods, which are potentially important
in application to molecular magnets but have not been actually used 
in the present study, are the availability of accurate forces in both codes,
the spin-orbit coupling treated within second order perturbation theory 
in NRLMOL \cite{PRB60-9566}, the possibility of general-form
(i.e., non-collinear) magnetization in {\sc Siesta} \cite{Fe-clus}.
Being an \emph{ab initio} pseudopotential code, {\sc Siesta} is usually
able to provide very good accuracy for the electronic structure 
and total energies without fitting to any exterior information, but still 
benchmark calculations 
make sense for critical cases and/or new systems. The present study
serves in part exactly this task, as the accuracy of all-electron
NRLMOL method is expected to be superior to that of {\sc Siesta}.
As the latter is routinely used for dynamical simulations in large systems
with virtually all symmetry lost, no special treatment is provided
for the use of possible point group symmetries available in the molecule.
For the systems in question, due to high degeneracy of energy levels
close to the chemical potential, the neglect of symmetry leads to
a certain instability in the convergence, which however can be managed
by keeping the mixing parameter very low (10$^{-3}$--10$^{-4}$)
and imposing the fixed spin moment scheme (FSM, Ref.~\cite{JPF14-L129}).

\section*{Results}

In the following, we discuss the results obtained by {\sc Siesta}
for the Li-centered molecule, and by NRLMOL -- for the Na-centered one,
with the coordinates as reported from the crystallographic study
of corresponding materials \cite{AnChIE36-2482}. The NRLMOL treatment
was restricted to the ground-state AFM configuration
(alternating orientations of Fe magnetic moments over the ring);
the {\sc Siesta} calculation addressed in addition different magnetic 
configurations, that allowed to make an estimate of exchange parameters. 

Fig.~2 displays the partial densities of states (DOS) on Fe
and its several neighbors in the AFM configuration, 
as calculated by both methods.
The discrete levels of the energy spectra are weighted (with the charge density
integrated over atom-centered spheres in NRLMOL, or according to Mulliken 
population analysis in {\sc Siesta}), and broadened for better visibility,
with broadening parameter of 0.15 eV  in {\sc Siesta} and 0.14 eV in NRLMOL. 
The local moments corresponding to integrating such partial DOS
over occupied states are given in Table \ref{tab:DOS}.

\begin{table}[b]
\caption{Local magnetic moments $M$ at Fe and its neighbors. NRLMOL
results correspond to spin density integrated over sphere of radius $R$
centered at corresponding atom; {\sc Siesta} values are due to Mulliken
popluation analysis.}
\begin{center}
\begin{tabular}{|l@{\hspace*{1.5cm}}r@{.}l
                  @{\hspace*{2.0cm}}r@{.}l
                  @{\hspace*{2.5cm}}r@{.}l|}
\hline
Atom & \multicolumn{2}{c}{\hspace*{-20mm}$R$(a.u.)} & 
       \multicolumn{2}{c}{\hspace*{-25mm}$M$($\mu_{\mbox{\tiny B}}$), 
                        NRLMOL} & 
       \multicolumn{2}{c|}{\hspace*{-10mm}$M$($\mu_{\mbox{\tiny B}}$), 
                        {\sc Siesta} } 
\\
\hline
Fe          & 2&19 &      3&85 &      3&91 \\
O (apical)  & 1&25 &      0&20 &      0&30 \\
O (bridge)  & 1&25 & $\pm$0&01 & $\pm$0&02 \\
N           & 1&32 &      0&07 &      0&09 \\
\hline
\end{tabular}
\end{center}
\label{tab:DOS}
\end{table}

One notes a remarkable agreement between both calculations up to
the finest details in the distribution of state densities at Fe and O
sites. This is the more astonishing as the local DOS is a loosely
defined property, normally dependent on the choice of region it attributes to, 
or, in the present context, to the choice of basis functions which are quite 
differently constructed in both calculational approaches. It also indicates 
that there is no significant influence of the central atom (Li or Na) 
on the electronic structure. Using NRLMOL we calculated the total charge 
within a sphere with a radius of 2.97 Bohr units around the Na atom. 
There are already 10.98 electrons contained inside the sphere and zero spin 
polarization which clearly indicates that the Na atom does not play a role for 
the electronic structure of the ferric wheel.

\medskip
\begin{figure}[h]
\centerline{\epsfig{figure=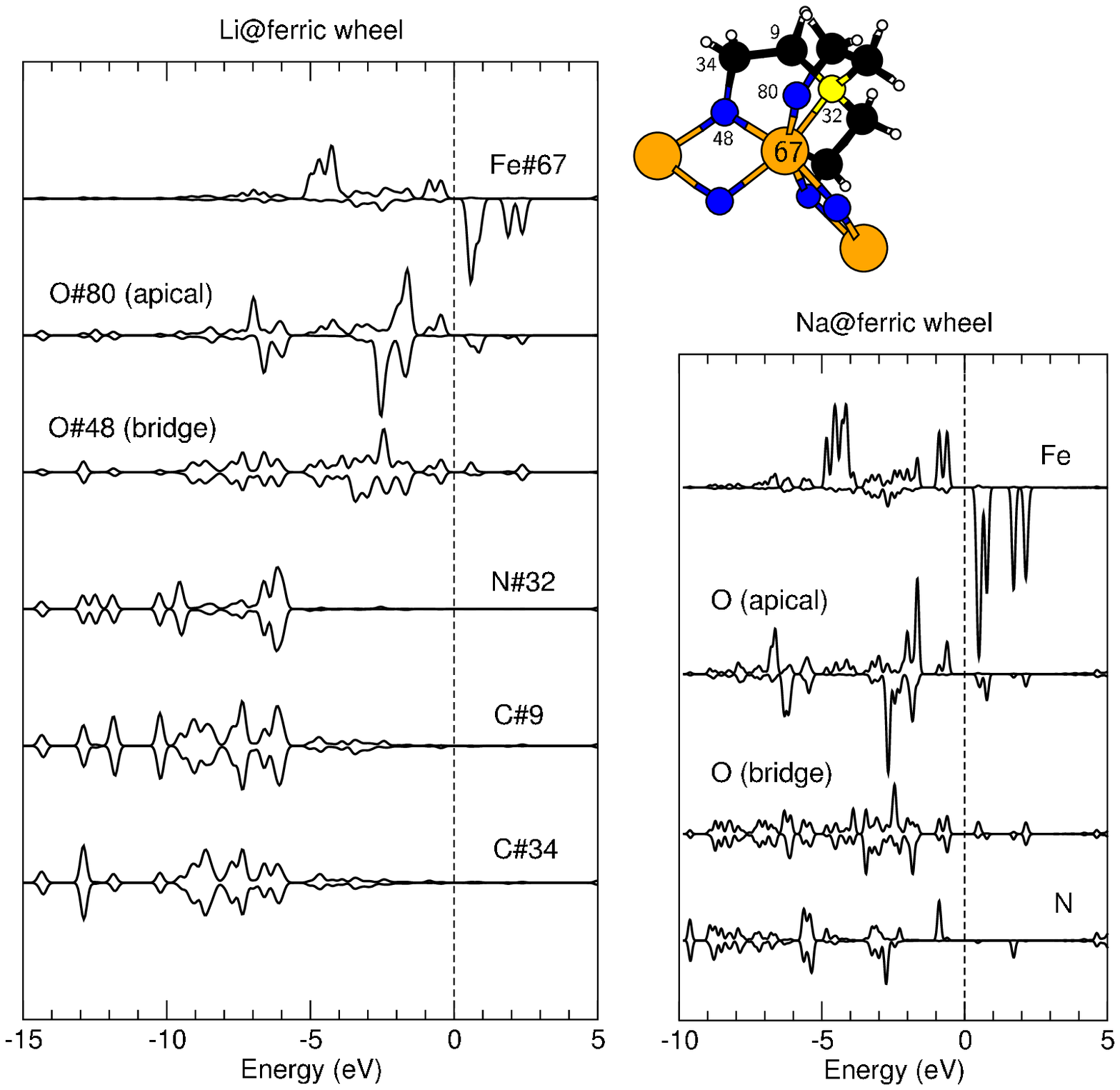,width=13.9cm}}
\caption{Atom- and spin-resolved partial densities of states
as calculated for Li-centered molecule by {\sc Siesta}
(left panel) and for Na-centered molecule by NRLMOL (right panel). 
The DOS at the Fe site is scaled down by a factor of 2 relative
to other constituents. The numbering of atoms which are neighbors
to the Fe atom is shown in the inset. See text for details of
calculation.}
\end{figure}

\newpage
Notably, both methods find the local magnetic moments on Fe sites 
are very close to 4 $\mu_{\mbox{\tiny B}}$
and \emph{not} to 5 $\mu_{\mbox{\tiny B}}$ as is generally
assumed, based on the above mentioned magnetization data. The maximal
magnetization $S$=5/2 of the Fe atom corresponds to a Fe(III)-ion with
in 3$d^5_{\uparrow}d^0_{\downarrow}$ configuration.
Our first-principles calculations suggest a somewhat different picture:
the minority-spin DOS has a non-zero occupation
due to the hybridization (chemical bonding) of Fe$3d$ with O$2p$ states.
However, the magnetic polarization in the organic ligand which provides the
octahedral coordination for the iron atoms, due to Fe is substantial, 
the most pronounced effect being on the apical oxygen atom 
(which is not participating in the bonding to the next Fe neighbor).
Taken together with the (smaller) polarization of the bridging oxygen atoms
and magnetization at the nitrogen site, the distributed
magnetic moment \emph{per} Fe atom yields 5 $\mu_{\mbox{\tiny B}}$,
recovering the agreement with the magnetization results.

An important consequence  is that the charge state
of iron is not Fe(III) but more close to Fe(II), according to our
calculations. Moreover, the distributed magnetic moment behaves like
a rigid one, in a sense that it can be inverted, following a spin flip
on a Fe site. This conclusion has been derived from our analysis of other 
magnetic configurations, done with {\sc Siesta} \cite{our_XPS}.
The local DOS does not change considerably when switching from 
AFM to FM configuration -- only the HOMO/LUMO gap becomes less pronounced,
and a slight ferro-magnetic shift between the two spin bands appears. 
Further, the change of total energies in the sequence of FSM, from the maximal
spin moment of 30 $\mu_{\mbox{\tiny B}}$ to one Fe spin inverted
(20 $\mu_{\mbox{\tiny B}}$), and then to two (second-neighbors) Fe spins
inverted is linear, supporting the conclusion that the system maps well 
onto the Heisenberg model. The corresponding exchange parameter, as estimated 
over both 30$\rightarrow$20 and 20$\rightarrow$10 $\mu_{\mbox{\tiny B}}$ flips,
is around $-$80 K. This yields a correct sign and correct order of
magnitude, relatively to experimental estimates of Ref.~\cite{InCh38-5879}.
However, the magnitude is overestimated by a factor of $\sim$4.
The exchange parameters $J$ depend on the spatial overlap of the
$d$-orbitals on different Fe-sites. It is well known that the $d$-orbitals
within DFT are not localized enough compared to experiment, consequently
the $J$ values will be overestimated. There are two main reasons for
this shortcoming. First, possible on-site correlations as known from
atomic physics are underestimated in case of ``conventional'' DFT.
Second, DFT is not free from spurious self-interactions due to the
replacement of the point-like electrons by corresponding densities.
Bringing in the atomic physics in the form of LDA+$U$ (adding a local orbital
dependent atomic Coulomb interaction parameter $U$ to DFT, \cite{LDA+U}) or 
self-interaction corrections (SIC) \cite{PRB37-9919} 
will improve the results by lowering the $d$-orbitals in energy and
therefore localizing them stronger. SIC only affects
occupied states, whereas LDA+$U$ plunges the occupied $d$-states and
shifts the unoccupied ones to higher energies.
By increasing, on the average, the magnetic excitation energy across
the spin majority-minority gap, both mechanisms
help to effectively reduce the magnitude of $J$.
In what regards the LDA+$U$ approach, this has been shown
by Boukhvalov \emph{et al.} for another molecular magnet,
Mn$_{12}$ \cite{PRB65-184435}.

\begin{figure}[h]
\centerline{\epsfig{figure=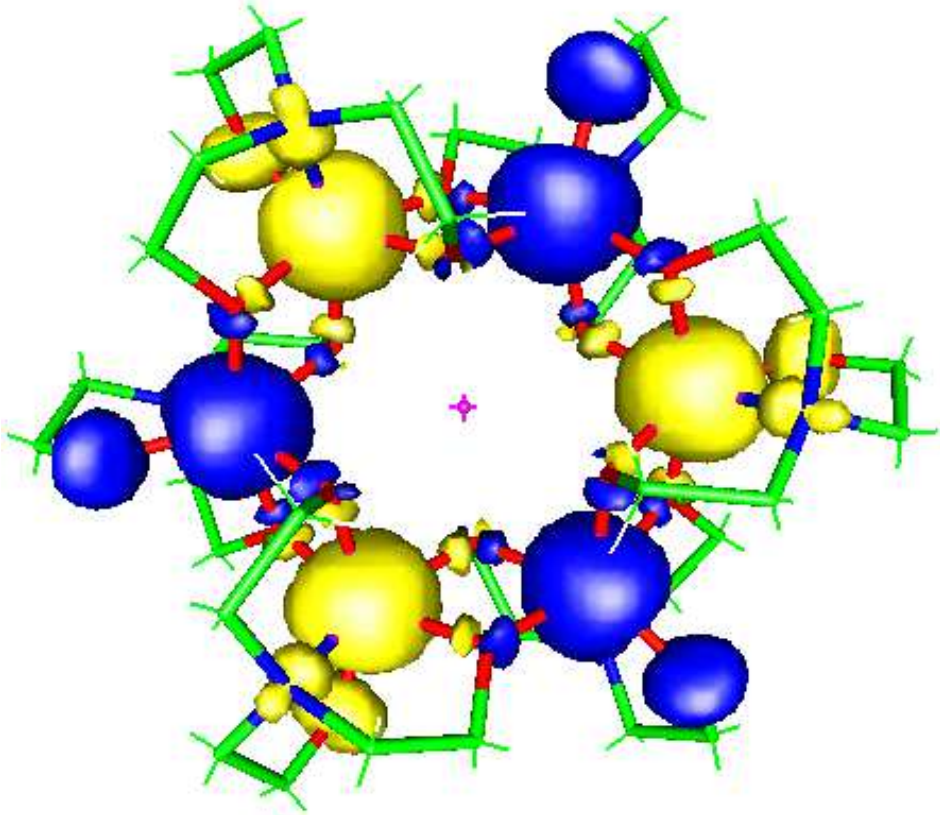,width=12.5cm,clip=true}}
\bigskip
\caption{Plot of the spin density map calculated with NRLMOL. 
The iso-surfaces correspond to $\pm 0.01 e/\mbox{\AA}^3$. While most 
of the magnetic moment is localized at the Fe atoms, there is still 
some spin polarization on O and N.}
\end{figure}

It is worth noting that the inversion of the third Fe spin
that yields the AFM configuration does not follow the same linear
trend. The origin of this is not clear at the moment and should be
tested in further high-precision calculations. 
We recently came aware of calculations by E.~Ruiz on similar systems 
\cite{Ruiz-private}, done also with the use of {\sc Siesta}.
According to them, the $J$ values extracted just from the 
FM--AFM total energy difference along the procedure described in
Ref.~\cite{JCompCh24-982} yielded the $J$ values of $\approx$ $-$24 K
(Li system) and $\approx$ $-$45 K (Na-system). A similar estimate from FM 
to AFM transition with our total energy data gives $-$38 K, i.e., 
such estimates seem to be in better agreement with the reported experimental 
data. It remains to be tested, however, whether the FM to AFM transition 
can be reasonably well fitted to the Heisenberg model with a single 
(nearest-neighbors) exchange parameter, in view of above mentioned
nonlinear dependence of the total energy on the total spin.
In V$_{15}$, another molecular magnet, a complex system of six exchange
parameters was found to be necessary \cite{PRL86-3400} in an otherwise 
conceptually similar fit to the Heisenberg model.

A clear visualization of the above discussed delocalized
(or, rather, distributed) magnetic moment associated
with the Fe atom comes from the map of spin density, obtained
from the NRLMOL calculation (Fig.~3). One should take into account that
the volume enclosed by the iso-surfaces is not directly 
correlated to the total moment at the site. 
One sees moreover an absence of magnetization on carbon and hydrogen
sites. The fact that the magnetization is noticeable and changes
its sign when passing through bridge oxygen atoms emphasizes 
the failure of methods depending the spherical averaging of
atom-centered potentials. 
{}From the side of experiment, it would be interesting to probe 
the spin polarization on oxygen and nitrogen atoms.

\section*{Conclusion}

We presented a study of the electronic structure of a six-membered ferric 
wheel with Li or Na as a central atom. The local densities of states 
on the iron and neighboring sites clearly indicate that the magnetic moment 
is distributed over several sites. The local moment at the iron site is still 
the largest one with about 4 $\mu_{\mbox{\tiny B}}$, although indicating that 
the iron in the ferric wheels is closer to Fe(II) than Fe(III) as generally 
expected. The local spin of $S$=5/2 per iron site as deduced from 
magnetization measurements is reproduced if one takes the moments 
on the neighbor atoms into account. The largest moment is found on the apical 
oxygen atom, followed by smaller moments on nitrogen and the bridging oxygen
atoms. This picture has been confirmed by a plot of the spin density map. 
A further goal of the present study has been a comparative
test of two different DFT implementations. In most cases the local 
(energy-resolved) DOS and (spatially resolved) magnetization density
show a very good quantitative agreement. The most pronounced deviations
are related to nitrogen, indicating less localized magnetic moment there
and thus higher sensitivity of numbers to the exact normalization
of DOS.

In order to explore the range of systems where the presented first-principles 
methods give reliable results further studies on more systems are required. 
In particular studies of experimentally well characterized classes of 
compounds where only the ligands are changed in a controlled way but not 
the magnetic core would reveal if our calculational approach can reproduce 
the observed changes in the magnetic behavior. Our current results make us 
very confident in the predictive power of the presented methods. This should 
allow for a microscopic understanding based on the electronic structure 
of single molecule magnets.

\section*{Acknowledgments}
The authors thank the Deutsche Forschungsgemeinschaft for
financial support (Priority Program ``Molecular Magnetism'')
and appreciate useful discussions with J\"urgen Schnack,
Oliver Waldmann, Sorin Chiuzb\u{a}ian, Manfred Neumann, Eliseo Ruiz
and Mikhail Katsnelson.
The crystal structure data have been kindly provided by the group of Prof.
Dr. R.~W.~Saalfrank from the Institute for Organic Chemistry of the
University Erlangen-N{\"u}rnberg.


\end{document}